\documentclass[aps,showpacs,preprintnumbers,amsmath, amssymb]{revtex4}

\usepackage{float}
\usepackage{graphics,epsfig}
\usepackage{graphicx}
\usepackage{dcolumn}
\usepackage{bm}

\begin{document}
\baselineskip=0.8 cm

\title{{\bf No scalar condensations outside reflecting stars with coupling terms from Ginzburg-Landau models }}
\author{Guohua Liu$^{1}$\footnote{liuguohua1234@163.com}}
\author{Yan Peng$^{2}$\footnote{yanpengphy@163.com}}
\affiliation{\\$^{1}$ School of Physics and Physical Engineering, Qufu Normal University, Qufu, Shandong, 273165, China}
\affiliation{\\$^{2}$ School of Mathematical Sciences, Qufu Normal University, Qufu, Shandong 273165, China}

\vspace*{0.2cm}
\begin{abstract}
\baselineskip=0.6 cm
\begin{center}
{\bf Abstract}
\end{center}

We consider static scalar fields coupled to the gradient
where the coupling also appears in next-to-leading
order Ginzburg-Landau models.
We study condensation behaviors of scalar fields outside
regular compact reflecting stars in the asymptotically flat background.
For non-negative coupling parameters, we prove that the
reflecting star cannot support coupled static scalar fields.

\end{abstract}

\maketitle
\newpage
\vspace*{0.2cm}

\section{Introduction}

The famous black hole no hair theorem has attracted lots
of interest from mathematicians and physicists.
It states that static scalar fields cannot exist outside
horizons of black holes in the asymptotically flat
background, for progress see references \cite{Bekenstein}-\cite{Brihaye}
and reviews \cite{Bekenstein-1,CAR}. It seems natural since
horizons can absorb scalar fields outside black holes.
However, such no scalar field property
also appears in horizonless spacetime.

In the horizonless asymptotically flat background,
Hod recently found that neutral compact reflecting stars
cannot support formation of scalar field hairs \cite{Hod-6}.
It was further found that such no hair property appears
for asymptotically flat neutral compact reflecting stars and scalar fields
nonminimally coupled to the curvature \cite{Hod-7}.
This no scalar hair theorem was also extended to
neutral compact reflecting stars in the asymptotically dS spacetime \cite{Bhattacharjee}.
It was further found that large charged compact reflecting stars
cannot allow the existence of scalar fields \cite{Hod-8}-\cite{Yan Peng-5}.
On the other side, in the extended Ginzburg-Landau superconductor model, there is
a new term with scalar fields coupled to the gradient \cite{cond1,cond2,cond3}.
So it is of some interest to generalize the discussion in \cite{Hod-6}
by considering scalar fields coupled to the gradient.

This work is organized as follows. We firstly introduce the
horizonless compact reflecting star gravity system with
exterior scalar fields coupled to the gradient.
For a non-negative coupling parameter, we prove that
horizonless compact reflecting star cannot support
the existence of static scalar field hairs. At last,
we present our main conclusions.

\section{No coupled scalar fields in the horizonless reflecting star background}

According to extended Ginzburg-Landau models, the scalar fields
may be coupled to the gradient \cite{cond1,cond2,cond3}.
We are interested in the case of coupled scalar fields
in the asymptotically flat spacetime, whose Lagrangian density
is given by \cite{dg,sh,Rogatko1}
\begin{eqnarray}\label{lagrange-1}
\mathcal{L}=R-|\nabla_{\alpha} \psi|^{2}-\xi\psi^{2}|\nabla_{\alpha} \psi|^{2}-V(\psi^{2}).
\end{eqnarray}
Here we label R as the Ricci curvature and $\psi=\psi(r)$ as the scalar field.
And $\xi$ is the parameter describing coupling strengthen between
scalar fields and the gradient. In order to obtain the final conclusion,
we assume $\xi>0$ in this work. The scalar field potential
$V(\psi^2)$ satisfying $V(0)=0$ and $\dot{V}=\frac{dV(\psi^2)}{d(\psi^2)}> 0$,
which is apparently satisfied by free scalar fields with $V(\psi^2)=\mu^2\psi^2$
and mass $\mu$.

The exterior spacetime of spherically symmetric horizonless compact star
is \cite{mr1,mr2,fc,Basu,Rogatko,Peng Wang}
\begin{eqnarray}\label{AdSBH}
ds^{2}&=&-ge^{-\chi}dt^{2}+\frac{dr^{2}}{g}+r^{2}(d\theta^2+sin^{2}\theta d\phi^{2}).
\end{eqnarray}
The metric functions $\chi$ and $g$ only depend on the
radial coordinate r. The angular coordinates are
labeled as $\theta$ and $\phi$ respectively.
And the radius of the compact star is labeled as $r_{s}$.

With the Lagrangian density (1), we obtain the equation of motion
\begin{eqnarray}\label{BHg}
(1+\xi\psi^2)\psi''+[(1+\xi\psi^2)(\frac{2}{r}-\frac{\chi'}{2}+\frac{g'}{g})+2\xi \psi \psi']\psi'-(\xi \psi'^{2}+\frac{\dot{V}}{g})\psi=0.
\end{eqnarray}

At the star radius, we impose the reflecting condition for the scalar field as
\begin{eqnarray}\label{BHg}
\psi(r_{s})=0.
\end{eqnarray}

At spatial infinity, metric functions have following asymptotical behaviors
\begin{eqnarray}\label{AdSBH}
\chi\rightarrow 0,~~~~~~g\rightarrow 1~~~~~~for~~~~~~r\rightarrow \infty.
\end{eqnarray}

The energy density can be expressed as
\begin{eqnarray}\label{BHg}
\rho=-T^{t}_{t}=g\psi'^{2}+g\xi\psi^{2}\psi'^{2}+V(\psi^{2}).
\end{eqnarray}

Physically acceptable solution requires that gravitational mass
$M=\int_{0}^{\infty}4\pi r^{2}\rho dr$ is finite, from which
we obtain the following relation
\begin{eqnarray}\label{BHg}
r^{2}\rho\rightarrow 0~~~~~for~~~~~r\rightarrow \infty.
\end{eqnarray}

At the infinity, relations (6), (7) and $\xi>0$ yields the condition
\begin{eqnarray}\label{InfBH}
&&\psi(\infty)=0.
\end{eqnarray}

According to relations (4) and (8), the scalar field $\psi(r)$ must possess at least
one extremum point ${r}_{peak}$, where the function reaches a positive local maximum
value or a negative local minimum value \cite{Hod-6}.
In the case that $\psi(r)$ reaches a positive local maximum value,
the scalar field satisfies the following relation
 \begin{eqnarray}\label{InfBH}
\{ \psi>0,~~~~\psi'=0~~~~and~~~~\psi''\leqslant0\}~~~~for~~~~r={r}_{peak}.
\end{eqnarray}

From relation (9) and $\xi>0$, we deduce the characterized inequality
\begin{eqnarray}\label{BHg}
(1+\xi\psi^2)\psi''+[(1+\xi\psi^2)(\frac{2}{r}-\frac{\chi'}{2}+\frac{g'}{g})+2\xi \psi \psi']\psi'-(\xi \psi'^{2}+\frac{\dot{V}}{g})\psi<0~~~~for~~~~r={r}_{peak}.
\end{eqnarray}

In another case that the scalar field reaches
a negative local minimum value, the relation becomes
\begin{eqnarray}\label{InfBH}
\{ \psi<0,~~~~\psi'=0~~~~and~~~~\psi''\geqslant0\}~~~~for~~~~r=r_{peak}.
\end{eqnarray}

From the relation (11) and $\xi>0$, we deduce the characteristic inequality
\begin{eqnarray}\label{BHg}
(1+\xi\psi^2)\psi''+[(1+\xi\psi^2)(\frac{2}{r}-\frac{\chi'}{2}+\frac{g'}{g})+2\xi \psi \psi']\psi'-(\xi \psi'^{2}+\frac{\dot{V}}{g})\psi>0~~~~for~~~~r=r_{peak}.
\end{eqnarray}

At the extremum point, the characteristic inequalities (10) and (12) are
in contradiction with the equation of motion (3).
It means the scalar field equation has no nonzero solution.
Then we obtain a conclusion that horizonless compact reflecting stars cannot
support static scalar fields coupled to the gradient
with non-negative coupling parameters.

\section{Conclusions}

We investigated the condensation behaviors of static
scalar fields outside asymptotically flat spherically symmetric
compact stars. At the star radius, we take the reflecting
boundary condition. In this work, we considered couplings
between scalar fields and the gradient, where this type of coupling also
appears in the next-to-leading order Ginzburg-Landau models.
In the case that coupling parameters are non-negative,
characteristic inequalities (10) and (12) are
in contradiction with the equation of motion (3),
which means that the reflecting star cannot support
static coupled scalar fields.

\begin{acknowledgments}

This work was supported by the Shandong Provincial Natural Science Foundation of China under Grant
No. ZR2018QA008. This work was supported by a grant from Qufu Normal University
of China under Grant No. xkjjc201906. This work was also supported by the Youth Innovations and Talents Project of Shandong
Provincial Colleges and Universities (Grant no. 201909118).

\end{acknowledgments}

\end{document}